\documentclass[12pt]{article}
\input{macros.tex}
\usepackage[export]{adjustbox}
\usepackage{rotating}
\usepackage{verbatim}
\usepackage{lscape}
\usepackage{graphics}
\usepackage{amsmath}
\usepackage{lscape}
\usepackage{supertabular}
\usepackage{hhline}
\usepackage{caption}
\usepackage{subcaption}
\usepackage{soul}
\usepackage[toc]{appendix}
\usepackage{tabulary}
\usepackage{etoolbox}
\usepackage{cancel}
\setstcolor{red}
\usepackage{ulem}
\usepackage{setspace}

\usepackage{booktabs}
\usepackage{multicol}
\usepackage{multirow}
\usepackage{siunitx}
\BeforeBeginEnvironment{appendices}{\clearpage}
\newcommand{\taci}[1]{}

\begin{document}

\thispagestyle{empty}
\baselineskip=27pt
\vskip 4mm
\begin{center} {\Large{\bf 
Towards physics-informed neural networks for landslide prediction}}
\end{center}

\baselineskip=12pt
\vskip 3mm

\begin{center}
\large
Ashok Dahal$^{1}$, Luigi Lombardo$^{1}$
\end{center}

\footnotetext[1]{
\baselineskip=10pt University of Twente, Faculty of Geo-Information Science and Earth Observation (ITC), PO Box 217, Enschede, AE 7500, Netherlands}

\baselineskip=16pt

\doublespacing
\begin{center}
{\large{\bf Abstract}}
\end{center}

For decades, solutions to regional scale landslide prediction have mostly relied on data-driven models, by definition, disconnected from the physics of the failure mechanism. The success and spread of such tools came from the ability to exploit proxy variables rather than explicit geotechnical ones, as the latter are prohibitive to acquire over broad landscapes. Our work implements a Physics Informed Neural Network (PINN) approach, thereby adding to a standard data-driven architecture, an intermediate constraint to solve for the permanent deformation typical of Newmark slope stability methods. This translates into a neural network tasked with explicitly retrieving geotechnical parameters from common proxy variables and then minimize a loss function with respect to the available coseismic landside inventory. 
The results are very promising, because our model not only produces excellent predictive performance in the form of standard susceptibility output, but in the process, also generates maps of the expected geotechnical properties at a regional scale.
Such architecture is therefore framed to tackle coseismic landslide prediction, something that, if confirmed in other studies, could open up towards PINN-based near-real-time predictions. To stimulate repeatability and reproducibility of the same experiment, we are openly sharing data and codes at the following GitHub repository: \href{https://github.com/ashokdahal/PINN.git}{https://github.com/ashokdahal/PINN.git}.

\baselineskip=10pt

\par\vfill\noindent
{\bf Keywords:} Physics-Informed Neural Network; Landslide Prediction; Regionalized Slope Stability; Deep Learning. 

\newpage
\baselineskip=16pt

\doublespacing
\section{Introduction}\label{intro7}
Landslides represent a major cascading geological hazard primarily triggered by two earth system processes: solid earth and hydrological processes.
Landslides induced by solid earth processes are typically caused by earthquakes, volcanic activity, and other geophysical phenomena, while the hydrological process predominantly triggers landslides through changes in groundwater levels and surface water saturation \citep{c2019a}.
In both instances, landslides occur due to increased stress, or ``driving force'', on a potential failure plane, with these forces being commonly referred to as triggering forces \citep{newmark1965effects,jibson1993,fan2019earthquake}.
The global distribution of landslides results in considerable loss of infrastructure and human lives \citep{petley2012global,fidan2024understanding}. 
Moreover, this trend is expected to escalate due to the increased frequency of hydrologically induced landslides under climate change scenarios \citep{gariano2016landslides,dahal2024junction} and heightened exposure due to development activities in areas susceptible to solid Earth-induced landslides \citep{reichenbach-2018,wang2024spatio}.
Therefore, the accurate and reliable modelling of landslides is a crucial and ongoing area of research.

In the case of earthquake-induced landslides, the slope potentially fails due to seismic loading in an otherwise stable slope \citep{gorum2015b}. 
Generally, earthquake-induced landslide hazard modelling involves understanding and estimating the required seismic loading conditions for a slope to fail \citep{Jibson2011}. This approach can be broadly categorized into two main types: physically based modelling and data-driven modelling \citep{fan2019earthquake}. Physically-based modelling solves the forward physical systems given the geotechnical properties of the material involved and the triggering force \citep{memon2018a}. 
In contrast, data-driven modelling employs statistical and machine learning-based models to simulate landslides, utilizing explicit and/or latent explanatory variables influencing landslide occurrence \citep{amato2023earthquake}.

Physics-based landslide modelling has been a well-established approach since the inception of soil mechanics \citep{terzaghi1950a}, and it is primarily based on force balance problems in physics. 
For earthquake-induced landslides, \citet{Jibson2011} categorizes these models into three main types: pseudo-static, stress deformation, and permanent deformation methods.

The pseudo-static method, as described by \citep{terzaghi1950a}, models seismic loading as a permanent body force acting on a static limit-equilibrium force diagram. When the force due to seismic loading exceeds the resisting force, the slope begins to fail. 
This intuitive and simplified method has a long history of use, but it only determines whether a slope is stable or failed without estimating the consequences of instability or the likelihood of failure \citep{Jibson2011}.

The stress deformation method is based on the finite element method developed by \citet{clough1990original}. 
This approach calculates stress changes due to reduced soil stiffness and resulting deformation due to seismic loading within a finite element mesh \citep{memon2018a}. 
Although this method is computationally intensive and requires detailed geotechnical parameters, it provides a naturally evolving failure plane and the most realistic representation of slope failure \citep{Jibson2011}.

The permanent displacement method, initially developed by \citet{newmark1965effects} and later simplified by \citet{jibson1993}, falls between the stress deformation and pseudo-static methods in terms of complexity and data requirements \citep{Jibson2011}. 
This method models a landslide as a rigid block on an inclined plane with a known critical seismic acceleration.
The slope begins to displace when seismic perturbation exceeds this critical acceleration. 
Once a certain displacement threshold is reached, the slope is considered to have failed. 
This method is fairly accurate if the slope geometry, soil properties, and ground motion are known \citep{jibson1993,hsieh2011a}.

All of the physics-based methods require spatially varying and detailed geotechnical parameters such as soil cohesion, internal friction angle, density, thickness, etc. \citep{fan2019earthquake}. 
This is possible for single or multiple slopes but is challenging for regional-scale research and applications. 

Data-driven methods for landslide modelling are mainly categorized into statistical and machine-learning approaches \citep{dahal2023explainable}.
Statistical methods determine landslides' probability, frequency or intensity based on environmental variables, which may be expressed 
explicitly or at a latent level \citep{lombardo2019geostatistical}.
Common statistical functions used in landslide modelling include bivariate models, logistic regression, generalized linear models, generalized additive models, and advanced models such as Cox processes \citep{atkinson1998generalised,brenning2008statistical,yalcin2008a,Steger2016,lombardo2020a,yadav2022joint}. 
Despite their differing mathematical formulations, these models share a common principle: identifying statistically significant relationships between independent and dependent variables. 
The dependent variables represent landslides' occurrence (susceptibility), frequency, or intensity (e.g., volume, area, impact pressure) \citep{nhess-24-823-2024,dahal2024junction}.
To define the statistical function that estimates landslides, statistical models are fitted using maximization or minimization techniques, aiming to produce outputs closely matching observations for a given set of input variables \citep{Taylor.Diggle.2014}.
Once fitted, the model's predictive capability is validated using an independent dataset to ensure its reliability for future estimations \citep{steger2016exploring}.

Machine learning methods follow the same principle as the statistical ones \citep{jackson1988a}. 

The main difference between the statistical and machine learning methods is that the letter do not assume a definite functional form, and the model has a potentially highly non-linear representation \citep{dahal2024junction}. 
Moreover, the optimization of deep learning techniques predominantly minimizes the loss function \citep{yang2019model}. 
The architecture and how the model operates are technical details that have been improved throughout the past decades, improving the given models' performance capability. 
The most common machine (and deep) learning models in landslide hazard modelling are artificial neural networks (ANN) \citep{gomez2005assessment}, recurrent neural networks (RNN) \citep{fang2023speechrecognition}, transformers \citep{dahal2024full}, and convolutional neural networks (CNN) \citep{nhess-24-823-2024}.

The main criticism of data-driven methods is that they do not respect the physical mechanism of the process they are modelling, and often, the neural networks cannot perform well outside of their calibration domain due to the excessive training on a specific data set or the unknown physics behind the retrieve functional dependence \citep{dahal2023explainable}.
Similarly, the physics-based approach is more difficult to constrain in the larger spatial domain due to the complexity of obtaining dynamic geotechnical parameters \citep{fan2019earthquake}. 
As such, the geotechnical properties of the material are also influenced by local climatological conditions such as prolonged precipitation or drought \citep{kramer1996geotechnical}.
Therefore, there is a need for a modelling framework for landslides that can infer the landslide given latent variables while respecting the process they are tasked to solve.
Currently, only two contributions have been published with a focus on physics-informed neural networks \cite[PINN;][]{liu2023physics,moeineddin2023physics}. 
This class of models is essentially placed in between physics and data-driven ones, with a neural network being the core solver, but tasked with learning the physics of the failure mechanism rather than ``blindly'' matching landslide presence/absence data to a set of predictors.
The first of the two contributions 
\citep{liu2023physics} makes use of the \textit{scoop3d} model \citep{reid2015a} to identify no-landslide (0 in Bernoulli distribution). 
This is how the authors justify their model to be physics-informed. 
This is somewhat a true statement, but also does not fully satisfies the requirement of a PINN, as per definition in physics or computer science \citep{raissi2019physics}.
In fact, the model proposed by \citet{liu2023physics} constrains the data in a physical manner, but the model does not formally incorporate any physical law. 
What the authors assume is that by generating landslide occurrence / non-occurrence locations through \textit{scoop3d}, then a neural network should learn the physics behind it. 
So, even if the attempt is certainly in the right direction, it still misses an important component in the standard definition of PINNs.

The second contribution in the topic was recently published by  \citet{moeineddin2023physics}. 
Their work integrates the creeping mechanism of landslides and correctly integrates the physics in the modelling framework. 
In this sense, they satisfy the PINN definition requirements. 
However, their approach is suitable for solving partial differential equations exclusively at specific locations. 
These are single slopes where geotechnical data is available. 
Therefore, the use of the PINN as per their proposed method, cannot be extended to other areas. 
In turn, this implies that the advantage of traditional data-driven models is still there, unless one defines a generalizable PINN. 

The aim of this manuscript is precisely to work towards a generalizable PINN, capable of respecting the physics of the landslide genesis, but also of being extended across large landscapes.
We will do so, by treating geotechnical parameters as the latent covariates of an architecture tasked with retreaving those as an intermediate step and make use of them to minimize the difference between landslide presence/absence locations and the estimated factor of safety. 

In the remainder of the manuscript, we present the data and area where we tested our PINN (Section \ref{sec:DataOverview}), followed by a methodological overview (Section \ref{sec:Methods}), which we wrote with a strong mathematical formalism to offer an engineering geological blueprint for those that may want to replicate the same model in the future. The manuscript then presents the results in Section \ref{sec:Results}, discusses them in Section \ref{sec:Discussion} and offers our vision for future developments in Section \ref{sec:Conclusions}.

\section{Data and test site overview}\label{sec:DataOverview}
\begin{figure}[ht]
    \centering
    \includegraphics[width=1\linewidth]{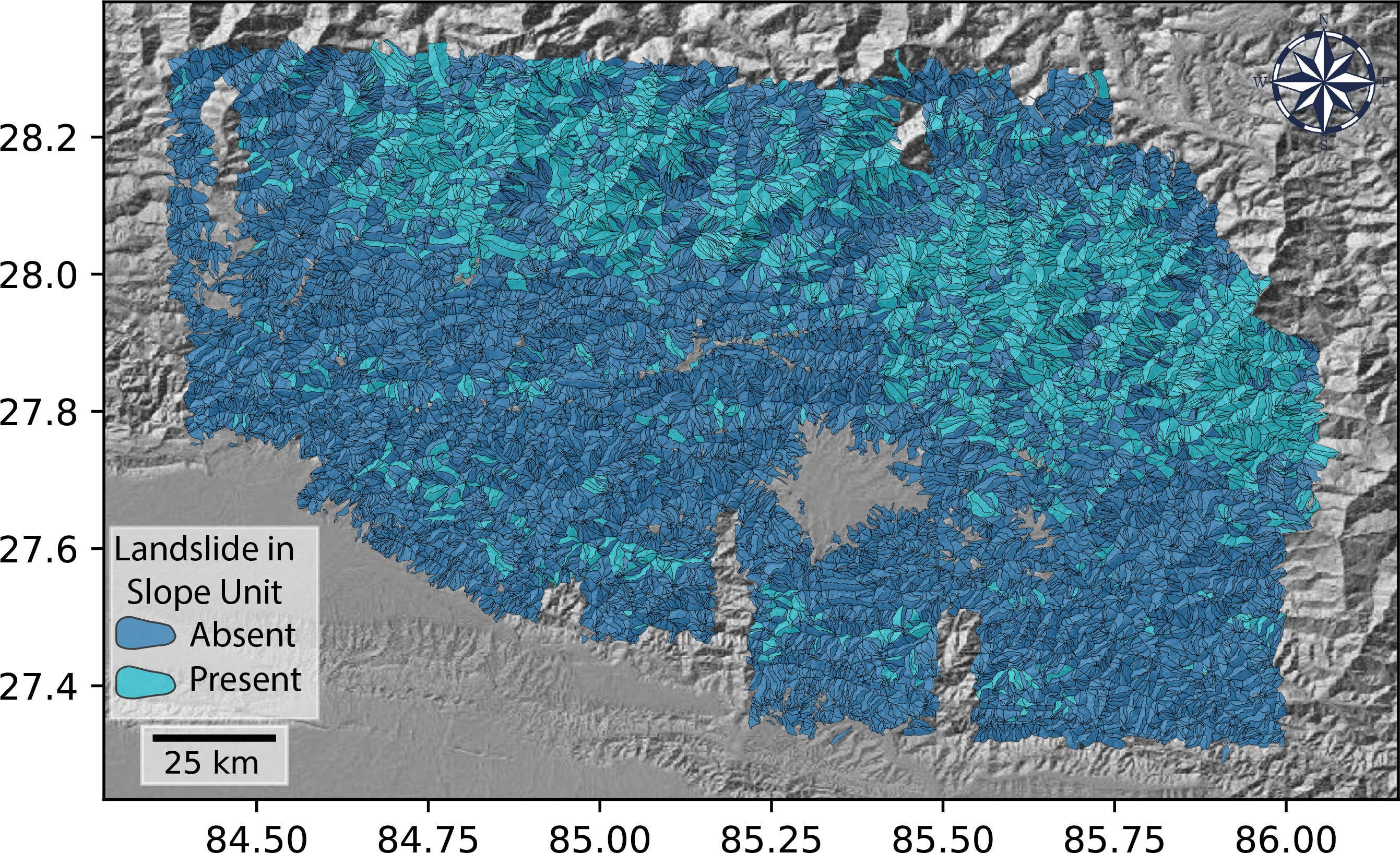}
    \caption{Study area showing the slope units with respective landslide presence and absence for 2015 Gorkha Earthquake according to \citet{roback2017map}}
    \label{fig:studyarea8}
\end{figure}

To train and evaluate our proposed PINN, we choose the same dataset freely distributed by \citet{dahal2023explainable}.
The study area was affected in 2015 by the 7.8 M$_w$ Gorkha earthquake, Nepal, resulting in tens of thousand coseismic landslides (see Figure~\ref{fig:studyarea8}). 
These have been mapped by \citet{roback2017map} in an openly accessible inventory.
Because of the combined use of manual mapping and very high-resolution optical images, the This inventory has been reported to be among the best available in the global repository made by \citep{schmitt2017open} and \citep{tanyacs2017presentation}, both in terms of quality \citep{tanyacs2019variation} and completeness \citep{hakan2020completeness}.

The Nepalese landscape where this study is conducted is essentially the roughest landscape on earth, due to the presence of the Himalayan range, being responsible for slopes ranging from 5$^\circ$ to 80$^\circ$ in steepness. 
This steepness is associated with  high mountain relief, hosting variations in elevation ranging from $\approx$ 500 meters to $\approx$ 5000 meters. 

Notably, because the Gorkha earthquake occurred during the Nepalese dry season, the contribution of the groundwater to the shallow failures (the dominant mechanism reported by Roback and co-authors) is assumed to play a negligible role \citep{regmi2016landslide}.  

In this experiment, most of the outcropping lithologies belong to the Siwalik formations, followed by the Himal group and various river formations such as the Seti and Sarung Khola formations \citep{dahal2012rainfall}.
The Siwalik Formation is primarily constituted by Molasse deposits of the Himalayas, comprising sandstones, mudstones, shales, and conglomerates \citep{upreti2001physiography}. 
The river formations, mainly found in the middle Himalayas, consist of a mixture of schist, granite, gneiss, phyllite, and quartzite \citep{upreti2001physiography,dahal2012rainfall}. 
Finally, the upper Himalayan region is where most of the tectonic compression is affected the lithological records, with metamorphic rocks encompassing schist, gneiss, migmatites, and marbles \citep{upreti2001physiography}. 
The geological map we used in this experiment was obtained from \citet{wandrey1998geologic}, thus partitioning the study area into nine different geological classes. 

Our mapping  unit of choice corresponds to slope units (SUs, hereafter). 
These are essentially half sub-basins considered to be homogenous in terms of their overall slope exposition  \citep{alvioli2016automatic}.   
Their use mainly appeared in lanslide studies as part of data-driven models \citep[e.g.,][]{tanyas2019global}, with only one example to date where they have supported physics-based slope stability models \citep{domenech2020preparing}, although exclusively during the post-processing phase. 

Our main objective is to make use of SUs to represent the ``infinite-slope'' problem \citep{xiao2016probabilistic,xi2024estimating}.

We recall here that for regional studies, SUs are typically considered
an ideal partition because they geographically approximate the hillslope response to a failure \citep{alvioli2018nation}.
Analogous considerations can be made in the framework of our proposed PINN. 
In fact, for a physics-informed data-driven approach, the same geomorphological considerations are valid, with the added geotechnical value of the homogeneity ensured by SUs being useful to express the inclined slope assumption predominantly held in pseudo-static and permanent deformation analyses \citep{newmark1965effects}. 

When defining our mapping units, we explored the potential use of the SU delineation made by \citep{alvioli2022geomorphological}. 
However, these are SUs generated to partition the whole Himalaya, thus making them too coarse for our modeling purposes.
For this reason, we opted to use the same delineation appeared in \citet{dahal2023explainable}. 
These had a much smaller initial target size, were selected among several possible versions generated through permutation of the \textit{r.slopeunits} paramerization \citet[see,][]{alvioli2016automatic} and ultimately validated through visual inspection.

Each of the SU served as the reference spatial scale at which we aggregated the input predictors as well as the landslide stable/unstable labels. 
For clarity, below we quickly summarize the reasoning behind the choice of our predictor set:

\renewcommand{\theenumi}{\roman{enumi}}%
\begin{enumerate}
    \item \textbf{Eastness:} The eastness shows how east direction the mountain slope faces. If it is facing completely towards the east, then the value will result in 1 and 0 for the contrary. It represents how much sunlight a particular slope gets, which can be a latent variable for landslide occurrence \citep{olaya2009basic}.
    \item \textbf{Northness:} Similar to eastness, northness provides information on the degree of northness of the mountain slope. This is also used for the exact same reason as eastness \citep{olaya2009basic}.
    \item \textbf{Horizontal Curvature:} The horizontal curvature provides on the overall horizontal ``bending'' of a given slope unit with respect to a imaginary horizontal tangent line. This information is a proxy for the 3-dimensional geometry of a sloped unit \citep{hengl2008geomorphometry}. 
    \item \textbf{Vertical Curvature:} Vertical curvature provides information on general vertical ``bending'' of the slope with respect to a given vertical plumbline \citep{hengl2008geomorphometry}.
    \item \textbf{Slope:} Slope provides information regarding the slope unit's steepness and is crucial information for any landslide hazard information as the acting gravity on a rigid body always acts with respect to the slope reference frame. Therefore, the slope is a first-degree control of the occurrence of landslides \citep{hengl2008geomorphometry,lombardo2019geostatistical}. 
    \item \textbf{Precipitation:}To inform the model about the relative wetness of the given slope, we included the total precipitation at a local slope unit for the three months before the main event. This information provides an overview of the relative wetness of the soil and failure surface \citep{moore1991digital}.
    \item \textbf{NDVI:} NDVI is a proxy for the root strength of vegetation present in the study area and it provides an overall quality and presence of the vegetation on a particular slope unit.
    \item \textbf{PGA:} The peak ground acceleration is another first-order control of the earthquake-induced landslides as it provides information on the perturbation to an otherwise stable slope due to the ground motion \citep{DAHAL2023108898}. 
    \item \textbf{Sand, Silt, and Clay Content:} This information provides the average quantity of sand, silt, and clay available on a given slope unit's top 2 meters of soil block. This information is important for a shallow landslide as properties of the material on the top 2 meters often represent the failure plane \citep{hengl2017soilgrids250m}. 
    \item \textbf{Bulk Density:} Similar to soil content, bulk density or the oven-dry weight of soil provides information on the material composition of the failure plane as well as the failure material\citep{hengl2017soilgrids250m}.
    \item \textbf{Geology:} Similar to ground motion and the slope, geology is another first-order control of the earthquake-induced landslides as it provides information on the overall type of geological structure of the slope unit and often has certain categories of the geological units have a higher correlation to landslide occurrence and even controlling the type of feasible landslides\citep{dahal2006geology,wandrey1998geologic}.
\end{enumerate}

\section{Methods}\label{sec:Methods}
Landslide susceptibility can be defined as the landslide occurrence probability $L(s,t)\in{0,1}$ in a specific location and time $L(s,t)=1$ \citep{fell2008guidelines,vanWesten2008}. 
When assessing how prone a given landscape is to generate slope failures, the use of historical \citep{maharaj1993landslide} or event-based \citep{lombardo2020chrono} landslide inventories often comes with neglecting the temporal component of the landslide genesis.  
Therefore, the susceptibility is spatially defined, $\mathcal{X}_s$, leading to a simplified notation as compared to the one presented above, $L(s)=1$. 

We recall once more that susceptibility is usually obtained by means of either physics-based or data-driven methods \citep{fan2019earthquake}. 
In the first case, the susceptibility $p_p(s)\in [0,1]$ is defined as the probability such that the displacement $D$ at any location $s$ exceeds a minimum displacement threshold $d_p$ (or velocity in some cases)
$p_p(s)={\rm Pr}({D(s)\geq d_m}\mid \boldsymbol{G}(s)=\boldsymbol{x}(s))$ for given geotechnical conditions $\boldsymbol{G}(s)$ \citep{huang2020a,huang2020integrated}.
In many cases, the variables $\boldsymbol{G}(s)$ cannot be gathered across regional sales, thus physics-based methods employed for such  scales simplify those variables by assuming constant geotechnical properties over the study domain $\mathcal{X}_s$ \citep{fan2019earthquake}. 

In this overarching theme, data-driven models offer valid alternative solutions. 
As statistical models can directly infer the probability of a random process conditioning it to a set of predictors, they directly estimate the probability of landslide occurrence $L(s)=1$ given the environmental conditions $\boldsymbol{X}(s)$, including topographic, geologic and other influencing factors \citep{dahal2024junction}. 
Given the dichotomous nature of $L(s)=1$, the Bernoulli probability distribution is well suited to address this classification task, with standard techniques denoting the probability distribution as $L(s)\sim {\rm Ber}(p_d(s))$. 
Therefore, the probability of landslide occurrence $p_d(s) \in [0,1]$ can be written as $p_d(s)={\rm Pr}\{L(s)=1\mid \boldsymbol{X}(s)=\boldsymbol{x}(s)\}$.

Given the two very different systems described above, the main problem with the data-driven method is that $\boldsymbol{X}(s)$ does not often represent the actual geotechnical conditions at which the slope failure occurs, and the model itself also does not respect the physical mechanism behind the landslide process \citep{dahal2023explainable}.
Therefore, the data-driven approach is ideally used as a first-order estimation of landslide hazard, from which the most susceptible locations can be selected and further analyses using physics-based approaches. However, there are two major issues in this sequence. 
Firstly, further analyses may miss hazardous locations if the data-driven approach misclassifies some highly susceptible locations, or is prone to type II errors. 
Secondly, such ideal implementation is useful only in a preemptive manner and cease to be practical during emergencies.
In situations where strong earthquakes induces widespread landsliding, what can be achieved is confined to a near-real-time landslide susceptibility assessment \citep{nowicki2018global}, and detailed analyses are prohibitive because of the urgency in disaster response \citep{mon2018analysis}. 

To solve these problems, we propose to combine the best of both worlds in a single model that offers regional scale predictions, geotechnical meaningful results and a near-real-time predisposition, once pre-trained. 

To do so, we estimate geotechnical parameters $\boldsymbol{G}(s)$ given environmental variables $\boldsymbol{X}(s)$ with some sort of highly non-linear functional approximation using deep learning function $\mathbb{F}$. Then, we use those parameters through a physics-based landslide estimation technique to produce a landslide susceptibility values. 
Therefore the approximated geotechnical parameters can be defined as $\boldsymbol{G}(s)$ = $\mathbb{F}(\boldsymbol{X}(s))\mid \boldsymbol{X}(s)=\boldsymbol{x}(s)$.

In this study, we opted to define a simple architecture where the neural network function $\mathbb{F}$ is a  ANN with 16 neural network blocks. Each block consists of a dense neural network of 64 neurons except for the last block, a batch normalization layer \citep{ioffe2015batch}, a dropout layer with a dropout ratio of 0.3 \citep{srivastava2014dropout}, and a rectified linear unit (ReLU) activation function \citep{yarotsky2017error,nair2010rectified}. 
The neural network itself contains the trainable weights and biases, which are initialized using random normal initialization \citep{thimm1995neural}. 
The batch normalization and dropout layers prevent the model from overfitting by regularizing the distribution of each batch and randomly deactivating 30\% of neurons. 
The ReLU activation changes the data flow to convert the model into a non-linear one. 
The last block of the neural network consists of a variable number of neurons, which are equal to the number of geotechnical parameters that we wish to estimate, which in this case is two.

Notably, which types and how many geotechnical parameters $\boldsymbol{G}(s)$ entails depends on our assumptions and the underlying physical mechanism we want to solve. 
Once $\boldsymbol{G}(s)$ is estimated with the function $\mathbb{F}$, it is then passed, together with required environmental conditions such as terrain slope, through a ``Landslide-Physics'' layer in the model, which respects the physics of the failure mechanism.

With this general framework in mind, we define our model as follows. 
First, we select a suitable physical mechanism for landslide estimation in earthquake-induced landslides. 
In this case, we opt for a simplified permanent deformation method (i.e., Newmarks' method)\citep{jibson2007,gallen2015coseismic}. 
This method is our choice of physical model for the ``Landslide-Physics'' layer. 
This layer defines landslide susceptibility as a function of permanent displacement due to seismic perturbation\citep{newmark1965effects}.
Let us assume that without any seismic loading, the slope has a ``standard'' factor of safety based on its geotechnical properties and geometry.
The geotechnical properties that affect slope stability are the cohesion of the material $C$, the thickness of the failure block $t$, the density of the rock and soil material $rho_r$, the density of water $\rho_w$, and internal friction angle $\varphi$. 
The safety factor is also largely controlled by the slope angle of failure plane $\alpha$, the ratio of saturated thickness to total thickness $m$, and gravitational acceleration $g$.
\citet{gallen2015coseismic,Gallen2017} defines the factor of safety based on those properties as:
\begin{equation}
\label{eqnfs}
F S=\frac{C}{t_m} \frac{1}{\left(\rho_r g\right) \sin (\alpha)}+\frac{\tan (\varphi)}{\tan (\alpha)}-\frac{m\left(\rho_w g\right) \tan (\varphi)}{\left(\rho_r g\right) \tan (\alpha)}
\end{equation}

Because the Gorkha earthquake occurred in the dry season in the Himalayan range, we assume that the water table was below the failure surface, rendering $m=0$ \citep{Gallen2017}.
This assumption simplifies the $FS$ equation. 
Moreover, following \citet{Gallen2017}, we assume the material's average and constant density over the study area with $\rho_r=2300 \si{kg.m^{-3}}$. 
The gravitational constant is $g=9.8 \si{m.s^{-2}}$. 
The factor of safety provides the slope's strength to resist the driving force; this can be converted to represent the minimum acceleration required to move the sliding block on an inclined plane. 
This acceleration term is called critical acceleration $a_c$ \citep{jibson1993}, formulated as:
\begin{equation}
\label{eqnac}
    a_c = (FS-1)g sin\alpha
\end{equation}

Now, by adding seismic loading in terms of peak ground acceleration $a_p$, we obtain an empirical model for the estimation of Newmark displacement of sliding block $D(s)$ \citep{jibson2007,gallen2015coseismic,Gallen2017}shown as:
\begin{equation}
\label{eqndn}
\log D(s)=0.215+\log \left[\left(1-\frac{a_c}{a_p}\right)^{2.341}\cdot\left({\frac{a_c}{a_p}}\right)^{-1.438}\right] \pm 0.51
\end{equation}

This whole process (equations, \ref{eqnfs},\ref{eqnac}, and \ref{eqndn} are combined in a layer representing function $\psi$ which takes the input of slope $\boldsymbol{\sigma}(s)\in\mathbf{X}(s)$ and peak ground acceleration $\mathbf{a_p}(s)\in\mathbf{X}(s)$ from environmental variables as well the estimated geotechnical parameters $\mathbf{G}(s)$. 
The geotechnical parameters are estimated using the neural network model $\mathbb{F}$ are the ratio of cohesion per thickness of failure mass $C/t_m$ and the internal friction angle $\varphi$.
The function $\psi$ returns the exponential of the permanent displacement term $\exp(D(s))$.
Mathematically, it can be represented as $D(s) = \psi(\mathbf{G}(s),\boldsymbol{\sigma}(s),\mathbf{a_p}(s))$. 
As we do not know the intermediate deformation of our landslide mass and we only know if the slope has failed or not, we define a commonly accepted threshold of 5 $\si{cm}$ as the displacement threshold above which the slope starts to move. 
However, this still does not provide the probability of failure (or susceptibility) but rather a dichotomous representation of the process. 
Therefore, we modified the ANN sigmoid function $\varsigma$ to be centered around 5 $\si{cm}$ threshold, such that it provides an physics-integrated data-driven susceptibility $p(s)\in[0,1]$ as:
\begin{equation}
\label{landslideactivation}
    p(s) = \varsigma(D(s)) =  \frac{1}{1+e^{(5-D(s))}}
\end{equation}

\begin{figure}[h]
    \centering
    \includegraphics[width=1.0\linewidth]{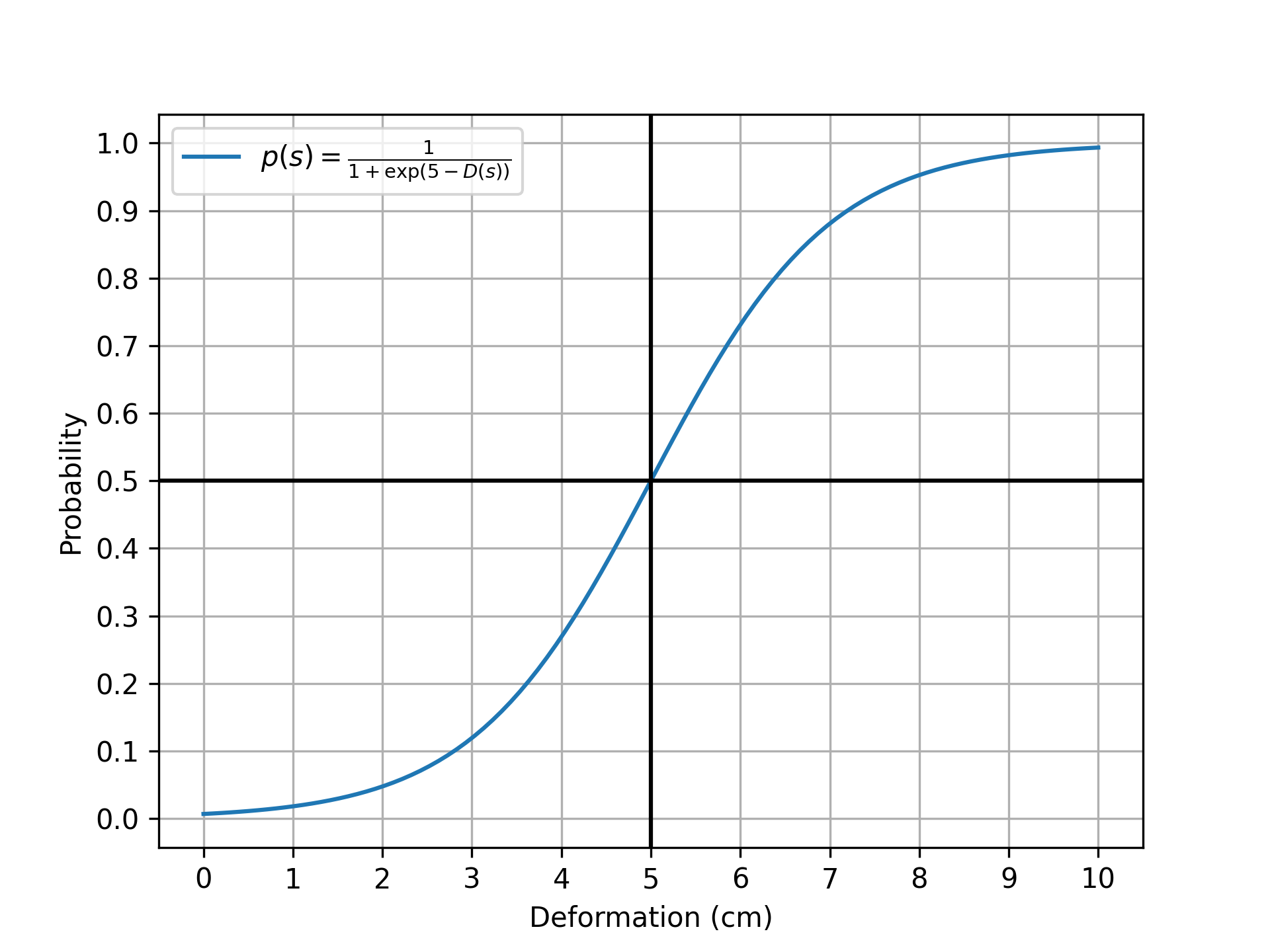}
    \caption{Figure representing the landslide activation function}
    \label{fig:landslideactivation}
\end{figure}

Specifically, equation \ref{landslideactivation} is similar to sigmoid activation with a centre around 5 cm and provides higher probabilities for the cases $D(s)\gg 5\si{cm}$ and lower probabilities for $D(s)\ll 5\si{cm}$. 
This function can be graphically represented in Figure~\ref{fig:landslideactivation}. 

This overall physics-integrated landslide susceptibility model can be written in the form $p(s) = \varsigma(\psi(\mathbb{F}(\mathbf{X}(s)),\boldsymbol{\sigma}(s),\mathbf{a_p}(s)))\mid \mathbf{X}(s) = \mathbf{x}(s), \boldsymbol{\sigma}(s) \in \mathbf{x}(s), \boldsymbol{a_p}(s) \in \mathbf{x}(s)$. 
This can be further simplified to denote the physics integrated model as $\Theta(\mathbf{X}(s))\mid\mathbf{X}(s)=\mathbf{x}(s)$.

We recall at this point that any neural network architecture relies on a loss function to converge to the best solution. 
In this sense, to train the trainable parts of the model $\Theta(\mathbf{X}(s))$, we can use a normal binary cross entropy loss function, because our model output is a pseudo-probability $p(s) \in [0,1]$ \citep{good1963maximum,akaike1998information}. 
To train our model, we have used a mini-batch optimisation technique where network parameters are updated with each randomly selected batch $B_N \subset {1,\ldots, n}$ of size $N<n$, where $n$ is the total number of batches \citep{li2014efficient}. 
Let the $\ell_i$ be the observed $L(s)$ value for each slope unit during the earthquake event, and $p_i(s_i)$ the predicted probability for $\text{i}^{\text{th}}$ variable.
This can be written as per equation \ref{binarycrossentropy}.

\begin{equation}
\label{binarycrossentropy}
\begin{split}
Loss = \bigg[-\sum_{i\in B_N} \left\{\ell_i \log(p_i(s_i))+(1-\ell_i)\log(1-p_i(s_i))\right\}
\bigg]
\end{split}
\end{equation}

With the above loss function and defined model, we trained our model using $70\%$ of the overall slope units. 
Out of the remaining $30\%$ of the dataset, half of it was used to validate the model, while the remaining half was used to test the it. 
The Adam optimizer \citep{https://doi.org/10.48550/arxiv.1412.6980} is selected for updating the model weights and biases (optimization) because of its robustness in training deep neural network models. 
We started the training process with $1e^{-3}$ learning rate, which was decreased by $90\%$ at each 10000 minibatch step. 
The batch size used for the training was 1024, meaning that 1024 randomly selected training datasets were passed to the model at each training step.
The model performance was tested using the area under the traditional receiver operating characteristics curve (ROC) curve as well as the balanced accuracy metric and F1-score \citep{Thibos1979,goutte2005probabilistic}. 
Moreover, to validate the usability of the model, we performed 10-fold cross-validation where nine near-equal random subsets of data are iteratively used to train the model while the complementary 10\% is used for testing.
Especially for binary classification models, a cross-validation procedure purely based on random data extraction is known to potentially produce overly optimistic results \citep{brenning2012spatial}. 
This happens because a random sampling essentially leave the spatial pattern in the data undisturbed, which in turn is translated in cross-validation performance results close to the ones obtained by fitting the model to the whole dataset. 
For this reason, the literature on the topic suggests to include spatial-cross validations, to complete the prediction performance overview \citep{pohjankukka2017estimating}. 
This is the case because a spatial cross-validation draws data not randomly but rather on the basis of their position, ensuring that any underlying spatial coherence is broken. 
This produces an unbiased overview of model performance. 

For the reasons above, we included a 10-fold spatial cross-validation as part of our analyses. 
In other words, we subdivided our SU dataset into ten spatial clusters, nine of which were iteratively selected for training, leaving the complentary one for testing. 
To visualize the ten SU clusters, see Figure 3 in \citet{dahal2023explainable}, as we used the same arrangement. 

Aside from the performance, the core of this contribution is in its geotechnical explainability. 
Therefore, it is important to showcase not only the probabilistic output $p(s)$ (i.e., the susceptibility map) but also an overview of the estimated parameters $\mathbf{G}(s)$, and how confidently our model retrieves their values per SU. 
In the way we defined our PINN, these parameters are the ratio of cohesion and landslide thickness as well as the internal friction angle. 
Additionally, we can also extract the critical acceleration ratio to understand how much seismic excitation is required to initiate a slope failure. 
To obtain the geotechnical parameters ($C/t_m$ and  $\varphi$) and their uncertainty, we bootstrapped the entire dataset 50 times by randomly picking 50\% of the data each time to train the model and test on the remaining dataset. 
We predicted the geotechnical parameters using each bootstrap model and estimated the median, $5^{th}$ percentile and $95^{th}$ percentile of the dataset. 
As a result, we can offer a statistical summary of the geotechnical parameter distribution for each SU.    

\section{Results}\label{sec:Results}
In this section, we present the results obtained from performance evaluation with spatial and random cross-validation, the intermediate geotechnical parameters, and landslide susceptibility. 

\begin{figure}
    \centering
    \includegraphics[width=0.65\linewidth]{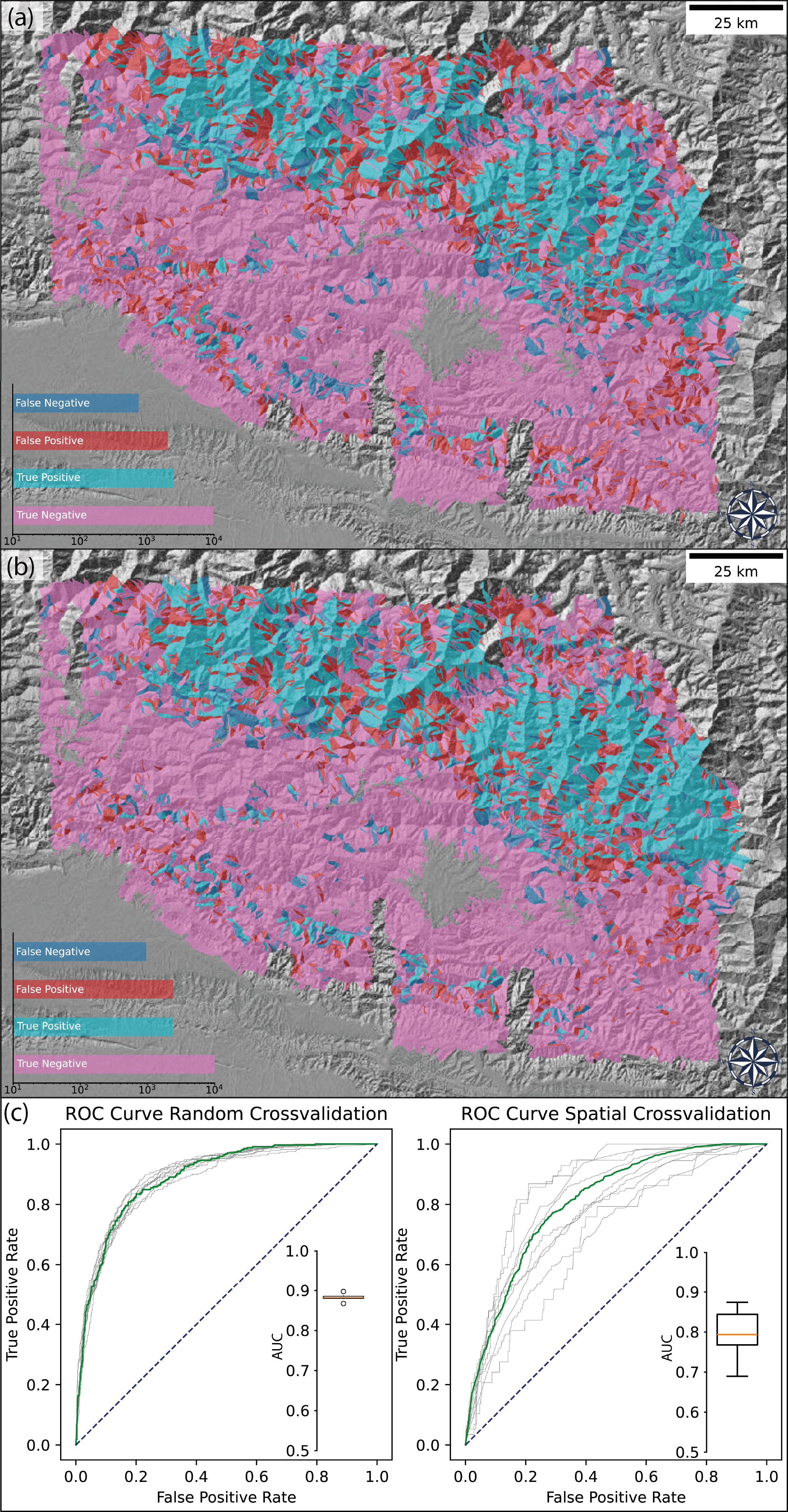}
    \caption{Performance evaluation of the defined model. (a) Confusion map of random cross-validation, (b) confusion map of spatial cross-validation, and (c) performance evaluation of receiver operating characteristics curve and their respective area under the curve.}
    \label{fig:products}
\end{figure}

In general, our PINN produced excellent performance scores \citep[see the classification proposed by][]{hosmer2000} with an area under the curve (AUC) of 0.87. 
To complete the performance evaluation, we also computed balanced accuracy and F1-score, these returning values of 0.79 and 0.78, respectively.
This shows that the reference model, built by using the whole dataset, can accurately differentiate stable and unstable SUs. Figure~\ref{fig:products} completes the performance assessment by presenting the two cross-validation schemes.  
The outcome of the random cross-validation metrics are shown in  Figure~\ref{fig:products}, with panel (a) presenting the result in a confusion map \citep{titti2022cloud} and panel (c) showing the corresponding ROC curves and AUC values. 
What stands out is that the model is able to predict most of the landslide locations in unseen datasets during 10-fold random cross-validation. 
Except for a few dark red and dark blue locations in the confusion map, most of the area shows a good fit with the observed dataset. 
Focusing on false positives and negatives, these are mostly present at the transition between SU clusters that failed and those that did not fail.
This translates in a near-outstanding classification performance, the K-fold random cross-validation producing consisten AUC values within the range of [0.86-0.90].

In an analogous manner, Figure~\ref{fig:products} also reports the confusion map produced through spatial cross-validation in panel (b), as well as the corresponding performance in panel (d).  

Even in this case, most of the misclassified SUs are present at the transition between failed and unfailed clusters of SU. 
Looking at the two confusion maps, if the overall patterns seem to agree, the spatial cross-validation produces more fluctuating performance, with AUC values ranging between [0.69-0.89].

Interestingly, because we used the same data and cross-validation structure as the one presented in \citep{dahal2023explainable}, we are able to compare the results between this contribution and the previous, based on a standard ANN.

When comparing the confusion map obtained through a random cross-validation, the present contribution and the previous seem to produce similar predictive patterns. 
It is important to recall that deep learning architectures are tasked with maximizing the predictive performance. 
Therefore, it is not surprising that a model not being informed by the underlying physics, can still perform well. 
In the end, this is why deep learning is so common nowadays. 

The most interesting aspect emerges when looking at the current confusion map and that of \citet[see Figure 3;][]{dahal2023explainable}.
In fact, the previous model, blind to any physical law, produced a large cluster of false positives in the south east of the study area. 
On the contrary, the map produce by our PINN does not present the same issue. 

Analogous considerations can be made examining the predictive performance, with the current spatial-cross validation producing a worst performance quantified in a 0.69 AUC, whereas the previous standard neural network returned a 0.58 for the same cross-validation fold.  

We would like to stress the importance of this last observations regarding standard and physics-informed neural networks, and refer the reader to  Section \ref{sec:Discussion} where we will further discuss this topic.


\begin{figure}
    \centering
    \includegraphics[width=1.0\linewidth]{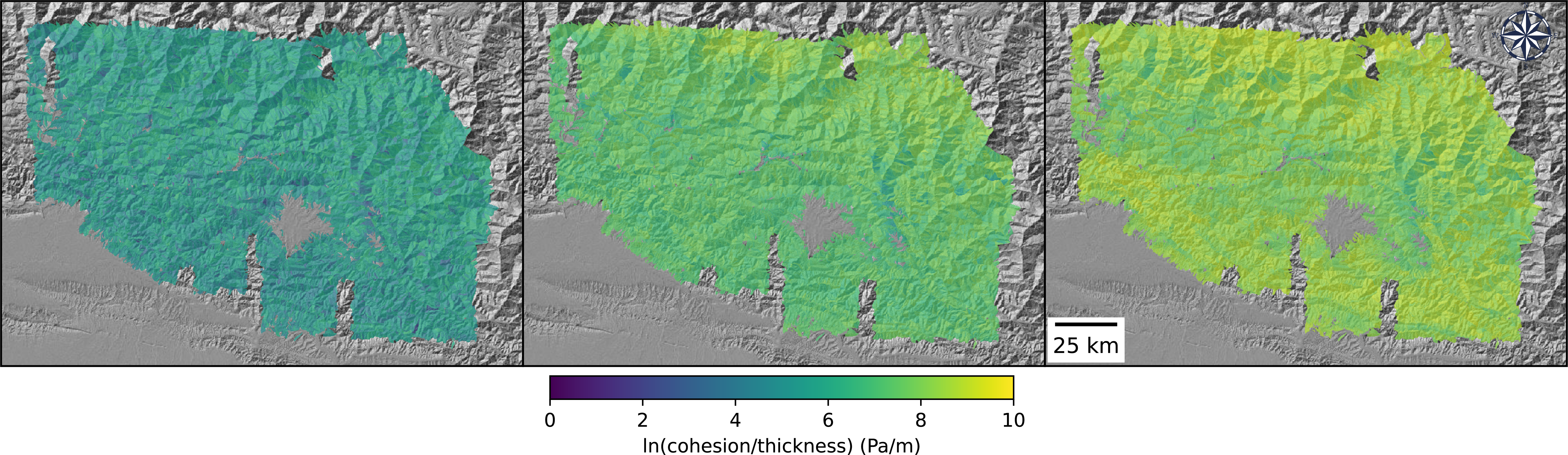}
    \caption{Cohesion per unit thickness of the failure body in log scale. The left subplot shows the $5^{th}$ quantile, the middle subplot shows the median value, and the right shows the $95^{th}$ quantile.}
    \label{fig:cohesion}
\end{figure}

\begin{figure}
    \centering
    \includegraphics[width=1\linewidth]{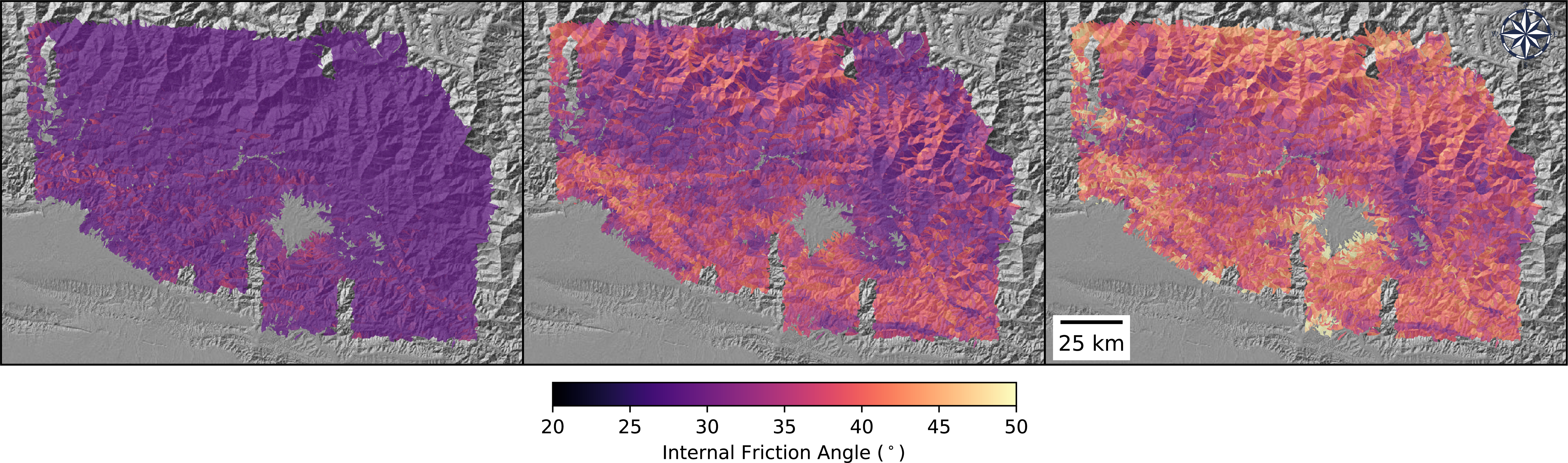}
    \caption{Internal friction angle. The left subplot shows the 5$^{th}$ quantile, the middle subplot shows the median value, and the right shows the 95$^{th}$ quantile.}
    \label{fig:phi}
\end{figure}

While data-driven models can often represent only performance evaluations and the final product, our framework also provides the capability to depict the estimated geotechnical parameters.
Here, we present the bootstrapped cohesion per unit thickness of the failure body $c/t$ and the internal friction angle $\varphi$ in Figures \ref{fig:cohesion} and \ref{fig:phi}, respectively. 
The values of $c/t$ and $\varphi$ appears to be saturated in both 5$^{th}$ and 95$^{th}$ percentile.
This is due to both percentiles being at the far ends of distribution, thus potentially being unsuited to represent the bulk of the distribution. 
Even though there are fewer spatial variations in those percentiles, the range of both parameters is within a commonly accepted and observed range of physical quantities \cite{Gallen2017}. 
This implies that the model is behaving according to the boundary conditions dictated by the physics of the process at hand. 
Looking carefully at the median, the $c/t$ has a higher range of values along the upper Himalayas. 
This corresponds to the northernost region of the study area, a sector mostly hosting rock materials. 
Conversely, $c/t$ values progressively decrease in the middle and lower Himalayas where the thickness of soil layer draping over bedrock materials becomes larger. 
Interestingly, some of the catchments show very low $c/t$ values, which might indirectly indicate a larger thickness of the landslide body, or a deeper potential sliding surface. 

Overall, the spatial patterns of $c/t$ appeared to be quite saturated, irrespective of the quantile of interest. This is likely because by taking the ratio of \textit{c} over \textit{t}, the latter term shrinks of the parameter distribution, allowing for a smaller spatial variability.
This is not the case for $\varphi$, as shown in Figure \ref{fig:phi}. 
There, a much higher spatial variability of the friction angle is retrieved, with reasonable patterns associated to the upper, middle and lower Himalayas. 
Specifically, the upper section of the Himalayan belt shows higher internal friction angles, wherease SUs where landslides occurred are associated to lower internal friction angles. 

\begin{figure}
    \centering
    \includegraphics[width=1\linewidth]{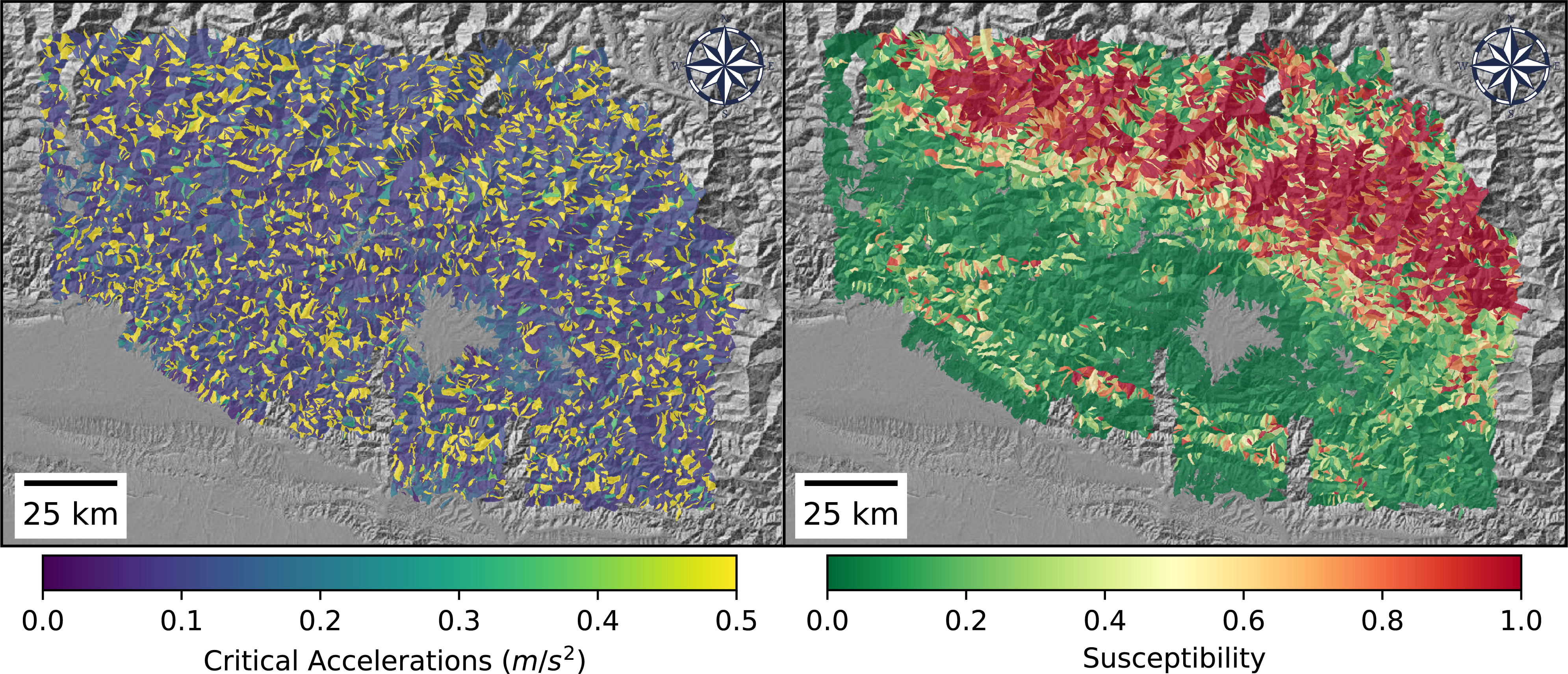}
    \caption{The obtained critical acceleration and susceptibility products. The left panel shows the critical acceleration and the right panel shows the susceptibility.}
    \label{fig:criticalaccnsus}
\end{figure}

To better understand the predicted results, we present the critical acceleration and the obtained susceptibility map through the main model in Figure~\ref{fig:criticalaccnsus}. 
The critical acceleration represents how much ground motion is required for a certain slope to fail, and it shows a mostly random distribution of high and low critical acceleration values, depending on their geotechnical parameters and slope angle. 
These slopes exhibit some level of spatial pattern that matches the low susceptibility SU, but aside from this, no visible pattern arises. 
Further discussions on the estimated critical acceleration will be provided in Section \ref{sec:Discussion}.
 
As for the susceptibility map itself, we can observe that the Middle Himalayan region where most of the landslides occurred are assigned with high susceptibility, and as we move southward, the susceptibility gradually decreases with few SU clusters appearing with higher landslide susceptibility.

\section{Discussion}\label{sec:Discussion}

The model we developed may constitute a valuable alternative for landslide hazard modelling, for it offers a series of advantages with respect to traditional data-driven solutions. 
The first element of strength, as the PINN definition implies, resides in the ability of our model to follow the physics of the coseismic failure process. 
Consequently, as the second element of strength, our PINN treats geotechnical properties as latent predictors, thus extracting them from data. 
On a more generic level, the weakness of many physically-based methods is precisely their need for explicit geotechnical information. 
In turn, they are either constrained to small geographic areas or potentially lead to lower prediction capabilities in large geographic areas. 
This is the case because acquiring geotechnical data is prohibitive when considering regional scales. 
Therefore, whenever applied to address regional slope stability problems, these models make assumptions that limit the variability of the required parameters. 
For instance, they assume a single value to be characteristic of the whole study area under consideration 
\citep{fan2019earthquake}. 

On the other side of the spectrum, data-driven models usually cannot work at the slope scale or at a scale involving few catchments. 
This happens because landslide inventories with numerous records are typically obtained over broader landscapes, unless some extreme trigger locally produced such conditions \citep[e.g.,][]{gorum2023preliminary,santangelo2023inventory}.
In other words, even if formally there is no minimum requirement in terms of landslide numbers to support any data-driven susceptibility assessment, it is also true that a dataset containing just tens of landslides cannot support the estimation of any meaningful statistical relation between stable/unstable slopes and the given predictor set.  

This is why data-driven susceptibility models are mostly successful when the involved geographic scales span over regional \citep[e.g.,][]{wan2009spatial,Frattini2010,lin2018framework}, national \citep[e.g.,][]{trigila2013landslide,lima2017landslide,wang2022space} and global \citep[e.g.,][]{jia2021global,felsberg2022estimating,tang2023global}  territories. 
However, despite the predictive power they ensure, physically-consistent results are not achieved \citep{glade2005landslide}. 
To date, the best one can do in this sense is to interpret the estimated regression coefficients \citep{steger2022deciphering,steger2024adopting} for statistical solutions or to interpret the predictor importance \citep{di2020machine,meena2022assessing} and more recently the SHAP \citep{Collini2022predicting,wang2024use} values for machine learning tools. 
 
We thus stress that our modelling approach consitutes a first step towards combining physically based and data-driven models to obtain physically consistent, generalizable and regional scale landslide susceptibility results.

Here, we have combined the permanent deformation approach proposed by \citet{newmark1965effects} into a deep learning model capable of estimating its parameters. 
Similar to adding ground motion and seismic loading to the pseudo-static method of \citet{terzaghi1950a}, such that the seismic perturbations may lead to a failure condition, our approach adds the physical constraint and seismic perturbations in data-driven landslide occurrence modelling.
Another problem our model tackles is that, even in the case where one can get access to a reliable geotechnical characterization of a regional landscape \citep[to our knowledge such example exists in Northridge, US;][]{DREYFUS201341}, the corresponding soil or rock sample are collected and tested in laboratory conditions.
This implies that the geotechnical characterization is valid for a specific location and time.
However, these may be subjected to changes in response to precipitation, vegetation, anthropic influence, etc. 
In turn, this requires physics-based models to be updated with new data or to be re-calibrated \citep{clarke2011quantifying,townsend2020quantifying,singeisen2022mechanisms}.
To some extent, our model is still affected by the same problem. 
However, the level of spatial details at which it retrieves the required geotechnical parameters corresponds to each individual SU, making it a tool that may serve the purpose of inferring a basic geotechnical characterization of a regional landscape. 
At an aggregated SU level, the geotechnical changes one may expect at the scale of a single sample should be less prominent. 
Therefore, our architecture should be, at least theoretically, less sensitive to changes, and if this issue would exist at all, it definitely provides a much cheaper way to generate a regional geotechnical characterization, as compared to a systematic field survey campaign.  

It is important to keep in mind that such advantages may come with limitations when comparing PINNs to standard deep learning architectures. 
In fact, PINNs add an additional and intermediate constraint in the form of physical laws. 
In general, this translates into lower prediction performance compared to a neural network who's free to converge to an optimal prediction function solely on the basis of the provided environmental variables. 
This is the generic assumption one can find in the technical literature on PINNs \citep[see,][]{cuomo2022scientific}.

Because we used the same data as \citet{dahal2023explainable}, we have the chance to use our previous work as a benchmark and verify this common hypothesis on PINNs versus pure performance-oriented architectures. 
Interestingly, in the present case and for both spatial and random cross-validation, we can see that the model performs similarly to the previous experiment, if not better.

For instance, we can observe that for the 10-fold random cross-validation case (Figure~\ref{fig:products}), the misclassified landslide locations are present at the boundaries between the SU with and without landslides. 

This might be because, due to the same environmental conditions, our model estimates similar geotechnical parameters.
Also, the peak ground motion data is spatially smooth.
Thus, the nearby SUs may appear to be exposed to similar ground motion characteristics.
This is a difficult issue to address due to the combination of the physically constrained approach and the available predictor data. 
Therefore, we observe that misclassification is mainly located near the transition between failed and stable slope units.

As for the case of spatial cross-validation, our PINN performed much better than the previous model without physics  \citep{dahal2023explainable}.
This is clearly visible in the southeastern sector of the study area, where our PINN essentially does not produce false predictions, whereas the previously published architecture erroneously classified the landscape.
We interpret these results in terms of spatial transferability. 
In other words, when data is insufficient or performance-oriented models predict over new data, our PINN provides more reliable estimates because they still need to obey the underlying laws of physics. 

Looking into the generated geotechnical characterization of the study area,  the estimated  $c/t$ and the $\varphi$ retrieve large spatial variations (in the median maps), as one would expect in nature or as opposed to generalized single-value assumptions.

The median value range shows a geotechnical valid range of parameters with a spatial variation where the higher Himalayan rock formations and the Siwalik regions have a higher $\varphi$, whereas the locations in the middle Himalayas with more dense soil composition have lower internal friction angle.
A similar observation can also be made with $c/t$, but as it is normalized to possible failure thickness, no distinct variation appears as shallow landslides usually occur in the low cohesion zone, and higher cohesive materials often have higher landslide depth.

As for the 5$^{th}$ percentile, both parameters present very small spatial variation, and the model parameters almost become constant, as constant values also can explain the landslide density as done by \citet{gallen2015coseismic}, this is still a realistic estimation.

In 95$^{th}$ percentile, we can observe that the $\varphi$ is saturated and has higher values throughout the study area with some exceptions and $c/t$ also has the same elevated value range.
This shows that the bootstrapped geotechnical properties show a reasonable value range in the median range and can be used as default, but we can also perform a full uncertainty propagation with all possible estimated values of geotechnical parameters.

The products obtained from the model for critical acceleration and landslide susceptibility have interesting properties.
The susceptibility product has a very similar spatial distribution to the landslide inventory we obtained.
In this case, the susceptibility range has a very distinct range for locations with and without landslides, which shows the model's confidence in estimating the landslide susceptibility.
One of the most interesting observations is that critical acceleration, being a function of geotechnical parameters and slope, has a random-looking distribution but shows similarities with slope units without landslides in regions where many landslides are present.
Critical acceleration is the amount of seismic excitation required for a slope to fail, and this can vary depending on the slope's properties. However, when the higher ground motion exceeds the critical acceleration, it eventually causes higher susceptibility and, eventually, higher landslide density. 

Using physics-integrated landslide hazard modelling could open up different possibilities for data-driven modelling as it is physically constrained.
This framework can be used to model landslides under climate extremes and different hazard scenarios similar to \citet{dahal2024junction}. 
Specifically, because AI-based data-driven models lack generalizability outside of the trained domain, but we have shown that integrating physics in the model can improve it. 
Therefore, one can use it both for rainfall-induced and earthquake-induced landslides in different climatological and geophysical scenarios \citep{cuomo2022scientific,maleki2022generalizability}. 
Moreover, having a mathematically and physically consistent model with the natural process can provide explainability, where one can explore intermediate products to understand if the model is behaving correctly or not. 
This human-on-the-loop approach is important for landslide susceptibility modelling in near real-time scenarios, as it is very important to have correct predictions to act on them \citep{dahal2023explainable}. 

Even though our model has provided a new framework for landslide hazard modelling, it still has some limitations and requires future research. 
The mathematical/physical framework that we used is based on permanent deformation analysis, and it is still a simplification of the process and assumes a consistent failure plane with a constant slope \citep{Jibson2011}. 
In nature, this is often unrealistic and requires simplification of slope by using slope units. 
Moreover, the failure plane is often assumed in this case.
The stress-deformation approach, however, can naturally generate a failure plane given the stress and strain scenarios \citep{griffiths1999slope}. 
However, it requires stress parameters and a refined mesh-based approach for landslide solution where each mesh element transfers its stress-strain properties to the next element \citep{memon2018a}. 
This is quite a common approach in physics-based modelling, but due to the lack of failure propagation on shallow landslide inventory, which occurs within very few seconds, this research could not use the stress-deformation approach.
However, it may be suitable for slow-moving landslides on a regional scale, where geotechnical parameters are estimated using a neural network, and then the solution for stress deformation is done via a finite-element-based approach \citep{moeineddin2023physics}.  

The landslides are often driven by a complex physical relationship between different earth system processes, and data-driven models are often incapable of modelling those processes sequentially or simultaneously \citep{fan2019earthquake}.
One of the main processes that control the landslide is a cascading hazard process, and it has many underlying physical mechanisms that can influence slope failures. 
This is mostly solved using the physics-based modelling approach \citep{Bout2021b,pudasaini2021mechanics}. However, it has the same limitations as we explained before and therefore, using a multi-process physics-based approach to solve landslide hazards would generate a more realistic cascading model.

Another limitation that we have faced in this work is that the ground motion data we have used comes from an empirical relationship \citep{wald2022shakemap}, and the permanent deformation model that we have used is also of an empirical nature \citep{jibson2007}. 
This is easier regarding modelling in near real-time, as the first ground motion estimation is available within an hour of the major earthquake.
We could further improve the quality of model estimation using full ground motion waveform as it fully respects \citet{newmark1965effects} approach and can be used in regional scales with full waveform ground motion simulations \citep{DAHAL2023108898}. 
The model structure that is required in this case would need to be changed from a simple artificial neural network model to a more complex recurrent \citep{medsker2001recurrent} or transformer \citep{jaderberg2015spatial} model.

\section{Concluding remarks}\label{sec:Conclusions}
Physics-informed modelling approaches are the future of scientific modelling in any system science, for they are robust and can combine the benefits of data-driven and physics-based tools.
In landslide science, data-driven and physics-based approaches have dominated two different and complementary aspects of hazard assessment procedures, with the former being the most sought solution at regional scales and the former being the standard for local ones.  
Bridging the gap between the two by combining their strengths can provide a valid alternative for landslide susceptibility and hazard modelling. 

Our modelling framework integrates the permanent deformation approach as part of a deep-learning architecture. 
This avenue uses latent variables to learn the geotechnical parameters and use them to estimate the physically constrained landslide occurrence probability. 
Our results show that the estimated values of geotechnical data lie in the observable and reasonable range, allowing for their variation over space. 
This leads to excellent model performance, reached while ensuring a clear physical explainability. 
In the immediate future, this framework can be extended as part of near-real-time prediction tools.

Even though current physical constraints fit the observed data well and show potential applications, this framework can be further improved by including a stress deformation approach in the landslide initiation. 
A better inventory with intermediate deformation and a stress-strain condition is required for this. 
Moreover, further improvement on this work could be made by including a multi-physical earth system process, which would provide a more general solution to the landslide susceptibility modelling. 

\clearpage\newpage
\bibliographystyle{CUP}
\bibliography{landslides}
\end{document}